\documentclass[utf8]{frontiersSCNS} 

\usepackage{url,hyperref,microtype,subcaption}
\usepackage[onehalfspacing]{setspace}
\usepackage{multirow}
\usepackage{subfloat}
\usepackage{rotating}
\usepackage{longtable}
\usepackage{tablefootnote}
\usepackage{chngcntr}

\usepackage[symbol]{footmisc}

\def\feii{{Fe {\sc ii}}}
\def\caii{{Ca {\sc ii}}}
\def\kms{km~s$^{-1}$}

\def\msun{M$_\odot$\/}
\def\oiii{{\sc [Oiii]}$\lambda$5007}
\def\rfe{R$_\mathrm{FeII}$}
\def\hb{H$\beta$}
\def\mbh{$M_{\rm BH}$\/}
\def\cc{cm$^{-3}$}
\def\civ{{C {\sc iv}}}
\def\nv{{N {\sc v}}}
\def\hei{{He {\sc i}}}
\def\heii{{He {\sc ii}}}
\def\ciii{{C {\sc iii}]}}
\def\lya{Ly$\alpha$}
\def\mgii{Mg {\sc ii}}
\def\oi{{O {\sc i}}}
\def\rblr{R$_{\rm BLR}$}
\def\lopt{L$_{\rm 5100}$}
\def\lbol{L$_{\rm bol}$}
\def\ledd{L$_{\rm Edd}$}
\def\lledd{$L/L_{\rm Edd}$}

\newenvironment{myfont}{\fontfamily{lmtt}\selectfont}{\par}
\DeclareTextFontCommand{\textmyfont}{\myfont}

\def\keyFont{\fontsize{8}{11}\helveticabold }
\def\firstAuthorLast{Panda \& Marziani}

\def\Authors{Swayamtrupta Panda\,$^{1,\dagger, \thanks{CNPq Fellow}}$ and Paola Marziani\,$^{2}$} 

\def\Address{$^{1}$Laborat\'orio Nacional de Astrof\'isica - MCTI, R. dos Estados Unidos, 154 - Na\c{c}\~oes, Itajub\'a - MG, 37504-364, Brazil\\
$^{2}$INAF-Astronomical Observatory of Padova, Vicolo dell'Osservatorio, 5, 35122 Padova PD, Italy
}

\def\corrAuthor{Swayamtrupta Panda}
\def\corrEmail{spanda@lna.br}

\begin{document}
\onecolumn
\firstpage{1}

\title[High Eddington AGNs: current state and challenges]{High Eddington quasars as discovery tools: current state and challenges}

\author[\firstAuthorLast ]{\Authors} 
\address{} 
\correspondance{} 

\extraAuth{}

\maketitle

\begin{abstract}

\section{}
A landmark of accretion processes in  active galactic nuclei (AGN) is  the  continuum originating from a  complex  structure, i.e. an accretion disk and  a corona around a supermassive black hole. Modelling the broad-band spectral energy distribution (SED) effectively  ionizing the gas-rich broad emission line region (BLR) is key to understanding the various radiative processes at play and their importance that eventually leads to the emission  from diverse physical conditions. Photoionization codes are a useful tool to investigate two aspects, the importance of the shape of the SED, and the physical conditions in the BLR. In this work, we critically review long-standing issues pertaining to the SED shape and the anisotropic continuum radiation from the central regions around the accreting supermassive black holes (few 10-100 gravitational radii), with a focus on black holes accreting at high rates, possibly much above the Eddington limit. The anisotropic emission is a direct consequence of the development of a geometrically and optically thick structure at regions very close to the black hole due to a marked increase in the accretion rates.
The analysis presented in this paper took advantage of the look at the diversity of the type-1 AGN provided by  the main sequence of quasars. The main sequence permitted us to assess the importance of the Eddington ratio and hence to locate the super Eddington sources in observational parameter space, as well as to constrain the distinctive physical conditions of their line-emitting BLR. This feat is posing the basis for the exploitation of quasars as cosmological distance indicators,  hopefully allowing us to use the fascinating super Eddington quasars up to unprecedented distances.


\tiny
 \keyFont{ \section{Keywords:} galaxies: active, quasars: emission lines; quasars: supermassive black holes; quasars: accretion, accretion disks; quasars: reverberation mapping; cosmology} 
\end{abstract}

\section{Active galactic nuclei as accreting black holes}\label{sec1}

Active galactic nuclei (AGNs) are among the brightest cosmic objects known to us \citep{weedman1976, weedman1977}. They harbour a supermassive black hole (SMBH) at their very centres which due to its immense gravitational potential allows for the infalling of matter. This in-falling matter loses angular momentum while being accreted onto the black hole. This accreted matter manifests in the form of a multi-colour accretion disk which gets heated up and radiates \citep{ss73,shields78,czerny87,panda18b}. The photon energy of the dissipated radiation spans a wide range of energies (from sub-eV to hundreds of eVs). The emitted photons then illuminate the material surrounding the accretion disk  and lead to the formation and emission of strong, broad emission lines \citep{schmidt63, greenstein_schmidt64, schmidt_green83, osterbrock_ferland06, netzer2015}. 


\section{Their Spectral Energy Distribution} 

AGN are observed over the entire range of the electromagnetic spectrum from the radio regime up to MeV-GeV-TeV energy $\gamma$-rays \citep{richards2006, harrison2014, yang_2022_cigale}. The classical view of AGN characterized by an almost flat spectral energy distribution (SED) over many decades in frequency -- to juxtapose to the ones of non-active galaxies -- has been 
   superseded and extended since long \citep{malkansargent82,malkan83},  by the recognition that the SED is different for AGN in different accretion states, and is most often characterized by significant features associated with diverse processes and aptly called bump or excess (i.e., IR excess, big blue bump, soft-X excess, etc.)\footnote{The old description of AGN continuum as non-thermal, featureless was perhaps  inspired by the earliest quasars studied that were mainly radio loud and at any rate sub-Eddington accretors. The power-law function used to fit the optical/UV continuum over a limited range in frequency is now considered to represent the thermal continuum from an accretion disk, whose power $P_\nu \propto v^\frac{1}{3}$\ \citep{ss73}, not as associated with a featureless synchrotron continuum. The synchrotron radiation from relativistic jets that accounts for most of the radio emission  is only a fraction of the optical continuum in non-blazar type-1  AGN or ``thermal" radio-loud AGN \citep[][and references therein]{antonucci12}. } 
   
   In this contribution, we shall restrict the attention mainly to the ionizing continuum in radio-quiet quasars. In this case, various components of the SED  arise due to different radiation mechanisms and at varying distances, notably among them: 
\begin{enumerate}
    \item  The characteristic `Big Blue Bump' \citep{czerny87, shields78} that is formed by the optical and ultraviolet radiation produced due to thermal emission from the accretion disk.
    \item The X-ray emission well-fit by a power-law, and produced when the UV photons from the disk undergo inverse Compton scattering by hot electrons in a Compton-thin corona close to the SMBH \citep[e.g.,][]{zdziarski1990}. 
    \item A spectral component observationally described as a ``soft X-ray excess'' \citep[e.g.,][]{arnaud1985}.  The most widely-accepted interpretation of the excess detected in soft X-rays is  of  emission in a Compton-thick  corona connected with the innermost accretion disk (\citealt{WF93,petruccietal20} and references therein). The competing model --  relativistically blurred photoionized disc reflection \citep{rossfabian,crummyetal16} --  is not anymore favoured as an explanation for the soft X-ray excess itself, although blurred accretion disk reflection can occur independently from the soft excess \citep{boissayetal16}. The soft excess helps bridge the absorption gap between the UV downturn and the soft X-ray upturn \citep[e.g.,][]{elvis94, laor97, richards2006, kubota18}, and changes the far-UV and soft-X-ray part of the spectrum, affecting the line production, including \feii\ emission in the BLR \citep{panda19a}. 
 \end{enumerate}

\begin{figure}[!htb]
\centering
\includegraphics[width=1\textwidth]{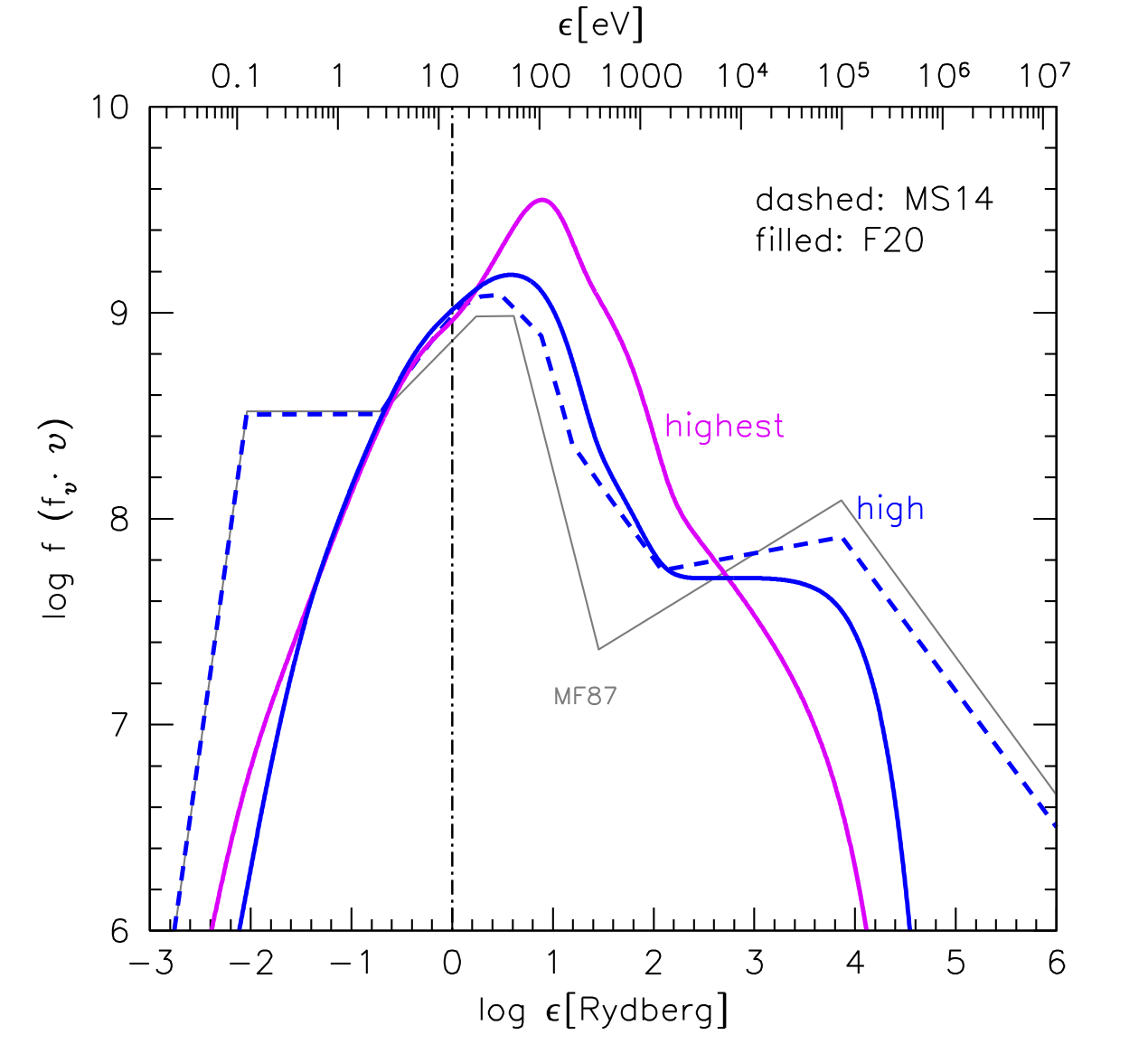}
\caption{
Templates of SEDs for high Eddington radiators. Grey; the landmark \citet{mathewsferland87}  SED; dashed blue: the \citet{marzianisulentic14} SED for quasars radiating close to the Eddington limit; blue and magenta: high and highest Eddington ratio templates from \citet{ferland2020}). The SEDs have been normalized at $\epsilon \approx $ 0.18 Ryd.
} 
\label{fig:sedax}
\end{figure}

 The ``intrinsic'' AGN continuum at photon energies high enough to ionize Hydrogen is therefore made of the thermal emission from the accretion disk, the power-law emission from the corona, and soft X-ray excess \citep{Collinson_2016PhDT.......352C, kubota18,panda19b, ferland2020}.   Figure \ref{fig:sedax} shows templates for quasars believed to radiate at  moderate or high Eddington ratio, $\eta \gtrsim 0.1 - 0.2$\ ({Population A and extreme Population A, hereafter xA, \citealt{marzianisulentic14} and \S \ref{popab}}), along with widely-exploited templates believed to be appropriate for populations of quasars radiating in this range \citep{mathewsferland87,marzianisulentic14,ferland2020}. There is a notable similarity between the curves. The SED defined by \citet{marzianisulentic14} for sources radiating close to the Eddington limit is in good agreement with the high case of \citet{ferland2020}. There is an increase in big blue bump prominence from the high to the highest case, the latter being associated with extreme values of the Eddington ratios. 
 {Note that the soft-X ray excess, located between the optical-UV bump and the peak at the hard X-ray ($\sim$100 keV), is prominent in between $\sim 1 $\ keV and 20 keV regions for the SED corresponding to the highest Eddington ratio case (magenta curve in Figure \ref{fig:sedax}) and marginally present in the high case (blue curve in Figure \ref{fig:sedax}). One notices that this feature steepens with $\Gamma > 2$\ as the Eddington ratio increases \citep{jin12b, ferland2020}. The feature almost disappears when one transitions to low Eddington ratio sources - see the blue dashed and grey SEDs, where the X-ray bump close to 100 keV is increasingly prominent.

 A weak but statistically significant correlation between hard-X photon index $\Gamma$\ and Eddington ratio has been found \citep{trakhtenbrotetal17,panagiotouwalter20,liuetal21}, and the statistical  weaknesses might be explained by the limited range of hard $\Gamma$\ values compared to uncertainties in individual $\Gamma$\ estimates  \citep{wangetal13}.  In the highest case, the soft and hard X-ray domains are very steep to the point that a turnover at $\sim 100$ keV as seen in Figure \ref{fig:sedax} for the  \citet{mathewsferland87} SED may not be anymore required. The difference between the \citet{mathewsferland87} SED and the extreme case of \citet{ferland2020} exemplifies this trend. The existence of X-ray weak type-1 AGN and their high prevalence among highly accreting sources \citep{zappacostaetal20,laurentietal22} may  also support the absence of a prominent Compton-thin coronal component in super-Eddington sources.} The sequence of SEDs in Figure \ref{fig:sedax} is related to the  4DE1 parameter space (Section \ref{ms}), although its connection to some of the parameters is still incomplete. 


A related issue is  the location of the  high energy downturn around $\sim$ 100 keV that is required by limits in the measured X-ray background \citep{mathewsferland87}. Observations are mostly available up to $\sim$ 20 keV, and  the energy of the downturn  is conventionally placed at $\approx 100 $\  keV in the SEDs of Figure \ref{fig:sedax} although it was not actually measured. In recent years measurements by NuSTAR and $\gamma$- ray observatories such as SWIFT indicate a dispersion in the actual turnover, from 50 to 200 keV \citep{fabianetal15,lubinskietal16}. It is currently debated whether the downturn energy may depend on the Eddington ratio, although the trend between $\Gamma$\ and the Eddington ratio suggests that a weak correlation might be possible \citep[][although see \citealt{molinaetal19}]{riccietal18}. However, we note that there are studies of multiple sources with cut-off energies measured by NuSTAR where the authors suggest that this cut-off energy is not dependent on the Eddington ratio or the black hole mass \citep[see e.g.,][]{2018A&A...614A..37T, 2022ApJ...927...42K}.


The focus of Figure \ref{fig:sedax} is for $\log \epsilon$ [Ryd]$\gtrsim -1$.  We mention in passing that in the NIR domain, as the low energy tail of the AD emission fades, extrinsic emission is  actually reprocessed  emission from the dusty torus which surrounds the accretion disk. It becomes the dominant emission at a few $\mu$m, along with the polar dust in the direction of SMBH spin axis \citep{netzer2015, padovani2017}. In the FIR, the SED might be dominated by dust heated by host galaxy star formation, more than by the AGN itself \citep{kirkpatricketal15}. This is occurring  in systems with high accretion rate \citep[e.g.,][]{marzianietal21}. In the case of highly accreting quasars, there is a relatively large prevalence of sources with high radio power, with radio-to-optical ratio $\gtrsim 1$\ \citep{gancietal19,wangetal22}, whose radio emission can be ascribed to star formation processes. Highly-accreting quasars might be predominantly seen as young  or rejuvenated active nuclei whose SED is  affected by star formation processes \citep[see e.g.,][]{caccianigaetal15,gancietal19}.

\section{Imminent challenges and opportunities}\label{sec2}

\subsection{What an AGN multi-frequency spectrum can reveal to us?}

From an observer's point of view, we can largely quantify the spectrum into two primary components: (1) the emission lines originating from the BLR/NLR clouds; and (2) the AGN  continuum, prominent beyond the Lyman limit, that can photoionize the surrounding  gas leading to line emission. The ionizing photon flux can be estimated by a careful analysis of the AGN SED, which then gives us a rough idea of the expected line fluxes for the multitude of ionic species (in their various ionization states) that we see in an AGN spectrum. A careful assessment of the density of these ionized clouds and their locations, in addition to the incident photon flux received by them,  allows us to {\em predict} the strengths of these lines. Important information about density, ionization conditions, and dynamics in the broad line-emitting region of AGN can be inferred from UV spectroscopic observations which are crucial to understanding these line-emitting regions. Past studies of highly-accreting quasars have {\em in turn } illustrated the use of certain line diagnostic ratios from observed spectra  (e.g., \civ/\heii, Al{\sc III}$\lambda$1860/Si{\sc III]}$\lambda$1892, \feii/\hb) in order to estimate these (density, ionization condition, and metallicity) parameters \citep[][and references therein]{negreteetal12, negrete2014, sniegowska+20, Garnica_2022}. Curiously, these extreme sources appear to be characterized by values of density, ionization, and metallicity that are extreme but also extremely well-defined: as far as the virialized emitting region is concerned the parameters reach   $n _\mathrm{H}\sim 10^{13}$ cm$^{-3}$, $\log U \sim -2.5$, $Z \gtrsim 20 Z_\odot$.\footnote{There is a general consensus that type-1 AGN BLR gas  has supersolar abundance with canonical estimates reaching over $10$\ times solar \citep{hamannferland93,hamannferland99}. However, the highest values depend also on the lines employed as diagnostics: the Al and Si lines are strongly dependent on enrichment by supernov\ae\ ejecta, as stressed by the authors themselves \citep[see e.g., ][]{Garnica_2022}}.      


\subsection{Dichotomy in optical and UV emission line profiles}
\label{dicho}

Historically, the BLR clouds were modelled as single clouds where the different lines arise from different parts of the same cloud - a picture that is still widely accepted \citep[][ see the BLR radial structure as shown by \citealt{negreteetal12}]{kwankrolik81}. In the mid-1980s, propositions were made to explain the BLR as two distinct components \citep{collin1988b, gaskell1982}. The broad emission spectrum in AGNs can be divided into two parts: the first set of lines that include \lya{}, \ciii{}, \civ{}, \hei{}, \heii{}, and \nv{} predominantly emitted by a highly ionized region that presumably has  a relatively low density ($\lesssim$10$^{10}$ \cc{}).   These are known as High Ionization Lines (HILs). The upper limit to the density of the media emitting these HILs is set by the semi-forbidden CIII] in order not to be collisionally de-excited even if the actual measurement of the blueshifted \ciii\ is problematic because of the blending with  Si{III}]$\lambda$1892. As a matter of fact, the density of the outflow is poorly constrained, and there is good reason to believe that a ``clumpy" scenario \citep[e.g.,][]{takeuchietal13}  might be also appropriate.  The second set of lines includes the bulk of the Balmer lines, \mgii{},  \feii{}, \oi{} and \caii{}, emitted by a mildly ionized medium having a much higher density ($\gtrsim$10$^{10}$ \cc{}). The real scenario is more convoluted and the search for a global unified picture is still ongoing. However, this representation  --- dichotomy into LILs and HILs  originating from the vicinity of the SMBH due to the inherent radiation of  the accretion disk -- has been instrumental to identify a  low-ionization virialized component and the contribution of a high ionization wind \citep{leighly04,marzianietal10}, that proved to be especially prominent in highly accreting sources i.e., quasar radiating at maximum radiative output per unit mass \citep{martinez-aldama2019, Panda_2022FrASS...950409P}. Low-ionization lines retain fairly symmetric profiles that indicate virial motions and therefore that their width is suitable for virial broadening estimation without the need of introducing large corrections \citep{marzianietal13,marzianietal19, Marziani_etal_2022}. \mbh\ estimates remain reliable even if the effect of partially resolved outflows (in radial velocity) has to be taken into account \citep{negreteetal18, Marziani_etal_2022, Buendia_Rios_2022arXiv220905526B}.



\subsection{Quasar Main Sequence}    
\label{ms}

\subsubsection{The Eigenvector 1 / Main Sequence}

The study of \citet{borosongreen1992} brought together the spectral diversity of Type-1 AGNs under a single framework. Their paper is fundamental for two reasons: (i) it provides one of the first templates for fitting the \feii{} pseudo-continuum. The \feii{} emission manifests itself as a pseudo-continuum owing to the many, blended multiplets over a wide wavelength range, extracted from the spectrum of a prototypical Narrow Line Seyfert Type-1 (NLS1) source, I Zw 1; and more importantly, (ii) it introduced the Eigenvector 1  (E1) sequence to unify the diverse group of AGNs. They used principal component analysis -- a conventional dimensionality reduction technique -- on observed properties of a sample of optically bright quasars to obtain a sequence. Of special importance is  the optical plane that shows the connection between the FWHM of the broad \hb{} and the strength of the \feii{} blend between 4434-4684 \AA\ to the \hb{} (or \rfe{}). This optical plane of the Eigenvector 1 (or of the ``main sequence" of quasars) was eventually included in a 4D parameter space (4DE1) that encompasses high-ionization broad line blueshifts \citep{sulenticetal00c,Sul2007}, and soft-X photon index \citep{sulenticetal00c,benschetal15}. The 4DE1 additional parameters are related to wind prominence and accretion status. It is therefore not unexpected that the main sequence might be primarily driven by the Eddington ratio i.e., the ratio between radiation and gravitational forces  \citep[e.g.,][]{sulentic2000, marzianietal01, boroson02, sh14, mar18} affecting several BLR physical properties \citep{panda18b, panda19a, panda19b, Panda_2021PhDT........22P}. 

\subsubsection{Population A and Population B}
\label{popab}

A classification based on the width of the \hb{} emission line profile in an AGN spectrum was introduced by \citet{sulentic2000}.  Population A includes local NLS1s as well as more massive high accretors which are mostly classified as radio-quiet \citep[e.g.,][]{marzianisulentic14} and that have FWHM(\hb{}) $\leq$ 4000 \kms{}. On the contrary, Population B sources are those with broader \hb{} ($\geq$ 4000 \kms{}), and are, at a large prevalence, ``jetted" sources \citep[e.g.,][]{padovani2017}. The Eigenvector 1 sequence of \citet{borosongreen1992} was extended to cover the soft-X ray domain \citep[e.g.,][]{sulenticetal00c}, assessing a relation between the soft-X photon index ($\Gamma_{\rm soft}$), the H$\beta$ line width, and the \feii{} prominence \citep[see also later developments on the relation between X-ray and optical spectra by ][]{grupe04,grupeetal10,aietal11,jin12b,benschetal15,ojhaetal20}.

Sources with higher values of soft X-ray excess (corresponding to a value of the soft-X photon index\footnote{here, the energy range considered for the estimation of the index is 0.5-2 keV based on archival Chandra and XMM-Newton data.} ($\Gamma_\mathrm{soft} \approx 3 - 4$) concentrate among the highly accreting Pop. A quasars \citep{grupe04,sulentic2008}, while Pop. B quasars typically have $\Gamma_\mathrm{soft}$\ $\approx$ 2.  The cut-off in the FWHM of \hb{} at 4000 \kms{} was suggested by \citet{sulentic2000} who found that low $z$\  AGN ($z \lesssim 1$) properties appear to change more significantly at this broader line-width cutoff (see also \citealt{collinetal06,ferland2020}). 

The usefulness of a fixed FWHM limit -- let it be 2000 \kms\ or 4000 \kms\ is questionable, as the FWHM is dependent on \mbh\ (or luminosity), viewing angle, and Eddington ratio \citep{mar18}. It makes sense if the limit is applied to samples in a narrow range of luminosity or \mbh. However, we reiterate the question posed a few years ago \citep{sul15}:

\begin{quote}
{\it Are populations A and B simply two extreme ends of the main sequence or do they represent two distinct quasar populations? Or are they tied via a smooth transition in the accretion mode?}    
\end{quote}

 The issue is very relevant to our quest to use quasars as standard, or standardizable candles, since the shape of the emission line profiles and continuum strength is directly connected to the central engine, especially to the black hole mass, and, the accretion rate, in addition to the black hole spin and, the angle at which the central engine is viewed by a distant observer \citep{wangetal14c, czerny2017, mar18, panda_frontiers, panda19b, panda_cafe2}. Looking at the  black hole mass vs luminosity diagram, type-1 AGN in optically selected surveys are distributed along a relatively narrow strip with $0.01 \lesssim$ \lledd $\lesssim 1$, over a  range of black hole masses \mbh\ exceeding 4 dexes \citep[see e.g., the Figure 15 of][]{donofrioetal21}. For \lledd$\lesssim$0.01, accretion is expected to enter into a radiatively inefficient domain, in addition to  selection effects that disfavour the lowest accretors at a given \mbh. At the other end of the \lledd\ range, sources that radiate at  \lledd$\gg$ 1 may simply not exist, as radiative efficiency is expected to decrease at a very high accretion rate, yielding to an asymptotic behaviour for the Eddington ratio toward a limiting value of order unity ($\sim 2 - 3$; \citealt{mineshigeetal00,wataraietal00,sadowski2011}) \footnote{Values of \lledd $\gg$ 1 should be viewed with extreme care, especially in cases of very narrow emission line profiles: orientation effects may drastically reduce the line profile in the case the emitting regions are seen pole-on. These sources stand out in a \lledd\ vs \mbh \ or luminosity diagram \citep{marzianietal06}.}. Perhaps the following scheme is already something more than a working hypothesis: Population B is associated with modest accretion rates, and the continuum may be fit by refined $\alpha$-disk models \citep{ss73,laornetzer89}. At some threshold of the radiative efficiency, $\eta \gtrsim 0.1$ the inner, advection-dominated  region of the disk starts to have a significant role in the geometry of the BLR. At a very high accretion rate, this effect may be extreme, with collimation of ionized outflows and shielding of the low-ionization emitting region from the luminous continuum that is instead seen by an observer oriented at a small angle with respect to the disk axis \citep[][Panda and Marziani in preparation]{wang14, Giustini_Proga_2019A&A...630A..94G}.
 It is also debatable to which extent accretion might be super-Eddington, as a large fraction of the infalling mass that would be accretion matter for the black hole might be actually expelled from the black hole gravitational sphere of influence \citep[e.g., ][]{carnianietal15,marzianietal16a,vietrietal18}.

\subsubsection{Narrow-line Seyfert 1s - a special class of AGNs?}

Narrow Line Seyfert Type-1 galaxies (or NLS1s)\footnote{Type 1/Type 2 classifications are based on the observation of the broad emission line features in an AGN spectrum. According to unified model \citep{antonucci93,urry_padovani_1995,marin14,netzer2015}, the presence of the dusty, obscuring torus impedes/allows the direct view to the central engine of the SMBH and the BLR region -- that is located closer to the SMBH. This then manifests in the AGN spectrum -- where the broad emission lines originating from the BLR are either seen (Type-1) or not (Type-2).} are a   class of Type-1 AGNs that are characterized with ``narrower'' broad emission lines:  FWHM(H$\beta_{\rm broad}$) $\leq$ 2,000 km s$^{-1}$, along with  the ratio of \oiii{} to the \hb{} less than 3 \citep{oster85, goodrich1989}. In addition to these properties, the NLS1s often exhibit strong \feii{} emission and the relative strength of the optical \feii{} (within 4434-4684 \AA) to the \hb{}, or \rfe{} $\gtrsim$ 1 \citep{sulentic2000, mar18, panda19b,rakshitetal20}. NLS1s have been used to analyze the \feii{} emission since the late 1970s \citep{phillips1978a} and have been regarded among the most noticeable cooling agents of the BLR, emitting about $\sim$25\% of the total energy in the BLR \citep{wills1985,marinelloetal16}. The \feii{} is a strong contaminant owing to a large number of emission lines and without proper modelling and subtraction, it may lead to a wrong description of the physical conditions in the BLR \citep{verner99, sigut2003, sigut2004, baldwin2004, panda_cafe2}. 
NLSy1s tend to be more variable than their ``broader'' counterparts in the X-ray regime \citep{grupe04, leighly1999, mchardy2006}, although the scales of their variability are not as pronounced in the optical and infrared regime \citep{giannuzzo1998, Ai2013}.

This is a summary of the conventional view of NLSy1s. A more exhaustive view is reached by the contextualization offered by the Eigenvector 1 Main Sequence. More prominently, the parameter \rfe{} is central to the E1 schema as it is the dominant variable in the principal component analysis presented by \citet{borosongreen1992}. The PCA analysis led for the first time  to the appreciation of the \feii\ relevance in large samples of quasar spectra.   NLSy1s showing significant \feii\ emission (\rfe $\gtrsim 0.5$) were described by \citet{sulentic2000} ``as drivers of all Eigenvector 1 correlations."  Indeed, NLS1s with high accretion rates are typically shown to have a soft-X-ray excess \citep{arnaud1985} in their broadband SED \citep{jin12a, jin12b, kubota18, ferland2020}.   NLS1s also show stronger blueshifts (blueward asymmetries) especially in the HILs \citep[e.g.,][]{leighly_moore2004, sulentic2000, Sul2007}, as well as higher \rfe\ implying high \lledd\ \citep[e.g.][]{duetal16}.  The E1 is now well understood to be associated with important parameters of the accretion process in the AGNs \citep{sulentic2000, sh14, mar18, panda19b, du2019, martinez-aldama_2021}, although the nature of the connection between \rfe\ and \lledd\ remains unclear to date.  NLS1s   typically host black holes with lower masses ($\lesssim$10$^7$ \msun{}) and tend to be less luminous and have low radio jet power --  which has led many authors \citep{sulentic2000,mathur00,fraix-burnet2017, Berton_2017} to link them to an evolutionary scheme of BHs. These authors have suggested that the NLS1s are the younger versions of more evolved, more massive SMBHs that  constitute the bulk of the population of AGNs.   

\subsection{Accretion parameters}

The estimation of black hole masses is perhaps the most sought-after analysis when it comes to AGN studies \citep{vester06,s11}. AGNs show variations in their continuum and emission line intensities that can range in the order of minutes/days for the continuum to days/weeks/months timescales for the BLR region \citep{ulrichetal97}. This crucial feature led to the estimation of black hole masses for a few hundred  nearby AGNs and relatively distant quasars\footnote{Quasars, or QSOs, are luminous AGNs discovered at larger redshifts. This distinction is mainly of historical importance; in the following, we will use the term quasars as an umbrella term for type 1 AGN.} using the technique of reverberation mapping \citep{blandford_mckee82, peterson2004, peterson88, peterson93} with the knowledge of the location of the line emitting region from the central SMBH\footnote{In actuality, the difference in the light travel time is estimated by making a cross-correlation between the continuum light directly reaching us and the light that bounces first at the BLR region and then comes to us \citep{peterson93,horneetal04}. The continuum light is produced very close to the SMBH --  in correspondence with  the inner  accretion disk \citep[see e.g.,][]{ss73, czerny87}.}.  Coupling the information of the velocity broadening from single/multi-epoch spectroscopy \citep{bentz13, dupu2014, kaspi2000} with a basic knowledge of the geometry of the emitting region \citep{Pancoast_2014,Li_etal_2016,lietal18}, we are well poised to derive the black hole masses using the virial relation \citep{petersonetal04}: \mbh $\sim f_\mathrm{S} r_\mathrm{BLR} \delta v^2 /G $.   There remains still a considerable level of uncertainty in the virial factor $f_\mathrm{S}$\ \citep{marzianisulentic12,shen_2013_BH}: the geometry and dynamics of the emitting regions remain poorly constrained. In dealing with \mbh\ estimate we actually encounter several difficulties associated with the lack of spherical symmetry of the BLR velocity field \citep{mclurejarvis02,decarlietal11}, and with the possibility of anisotropic continuum emission. The uncertainties in the \mbh\ estimation affect the \lledd\ estimates as well, with the complication that \lledd\ estimates from {optical/UV} data depend on a bolometric correction that in turn depends not only on  accretion state as evinced from Figure \ref{fig:sedax} but also on luminosity \citep[e.g.,][and references therein]{Netzer2019} as well as on viewing angle \citep[e.g.,][]{Runnoe_etal_2013}. Curiously, if the assumption of a highly flattened \hb\ emitting region is correct, there could be a way out right for super-Eddington sources exploiting the computation of the ``virial luminosity''  from line widths (Section \ref{cosmoxa} and Eq. \ref{eq:lvir}). 

The ratios between the virial and the redshift-based luminosity can be explained entirely by orientation effects \citep{negreteetal18}, thereby making it possible to derive an estimate of the viewing angle for each individual source. This method has been barely explored, but it has in principle the ability to constrain the disk/wind scenario in extreme Population A.  We will discuss in \S \ref{sec3} the anisotropy in continuum emission for xA sources - these refer to the sources that exhibit strong \feii{} emission (\rfe{}$\gtrsim$1) and are associated with accretion rates close to the Eddington limit, in the optical plane of the Eigenvector 1 main sequence diagram \citep{marzianietal18, panda19b}.  



\section{An avenue for cosmological studies?}
\label{cosmoxa}

A better understanding of the AGN's inner workings  can pave the road to far-reaching applications. One of them is the standardization of quasars (or QSOs) for measuring cosmological parameters. Two methods that involve quasar intrinsic properties resort to a law analogous to that of Faber-Jackson, connecting velocity dispersion and luminosity (Section \ref{xaedd}), and the radius-luminosity scaling laws (Section \ref{rl}). Both methods face challenges. 

\subsection{Eddington standard candles?}
\label{xaedd}

The application of quasars radiating at or above the Eddington limit has been proposed for several years although the method has not yet been  exploited to its full potential \cite[][and references therein]{marzianietal21,dultzinetal20}. The method is conceptually simple: the accretion luminosity of a quasar is proportional to a power of the line width\footnote{The equation is equivalent to the original formulation of the Faber-Jackson law \citep{faberjackson76} and is equivalent to other relations linking virialized systems to the amount of radiation emitted.}, i.e., $L \propto$FWHM$^\mathrm{n}.$ The value of the exponent $n = 4$ comes from the virial relation for the black hole mass, the assumptions of constant Eddington ratio (\lledd \ $ \propto L/$\mbh $\approx const.$), and of BLR radius rigorously scaling   with luminosity as $r \propto L^{0.5}$. The last assumption is likely to be verified for sources radiating close to the Eddington limit: they are identified by spectral similarity (\rfe $> 1$), and so SED and  the physical properties of the emitting regions  need to be similar. 

More in detail, the equation connecting luminosity and line width can be written as: 
\begin{equation}    
L \sim {\mathcal L_\bullet} \eta^2 f_\mathrm{S}^2(\theta) \mathcal{S}(\mathrm{SED}) \frac{1}{\mathcal{P}}\delta v_\mathrm{r}^n \label{eq:lvir}
\end{equation}

where we have evidenced the main physical factors entering the estimate of the virial luminosity from  measurements of radial velocity dispersion  $\delta v_\mathrm{r}$\ (FWHM, $\sigma$).  The   effect of orientation can be quantified  by assuming that the line broadening is due to an isotropic component $\delta v_\mathrm{iso}$ + a flattened component whose virial velocity field projection along the line of sight is 


\begin{equation}
\delta v^{2}_\mathrm{r} = \frac{\delta v_\mathrm{iso}^{2}}{3} + \delta v_\mathrm{K}^{2} \sin^{2}\theta.
\label{eq:v}
\end{equation} 
that implies:
\begin{equation}
f_\mathrm{S}(\theta) = \frac{1}{\frac{1}{3} \left(\frac{ \delta v_\mathrm{iso}}{\delta v_\mathrm{K}}\right)^{2} +\sin^{2}\theta}
\label{eq:v}
\end{equation} 

The factor $\mathcal{S}$\ is the ratio between the SED fraction of the ionizing continuum and the average energy of the ionizing photons. The factor $\mathcal{P}$ is the product density times ionization parameter ($n_{\rm H}U \sim 10^{9.6} $cm$^{-3}, $ \citealt{padovanirafanelli88, matsuokaetal08,negreteetal12}). The two factors come from the definition of the photoionization radius of the BLR \citep{wandel99}. 
Deviations between the virial estimates and luminosity estimated from redshift and assumed concordance cosmology can be fully explained by the effect of orientation \citep{negreteetal18}. The distributions of the viewing angles from the \cite{negreteetal18} sample based on \hb\ at low $z$ peaks at about 17 degrees, with only a very small fraction of quasars, is observed at $\theta \gtrsim 30^{\circ}$. This means that the effect of kinematic anisotropy on the computation of the virial luminosity should introduce a significant dispersion, as it is $\propto 1/\sin^2\theta$.

\subsection{Scatter in the R-L relation, standardizing QSOs for cosmological studies \label{rl}}
    
\begin{figure}[!htb]
    \centering
    \includegraphics[width=0.495\textwidth]{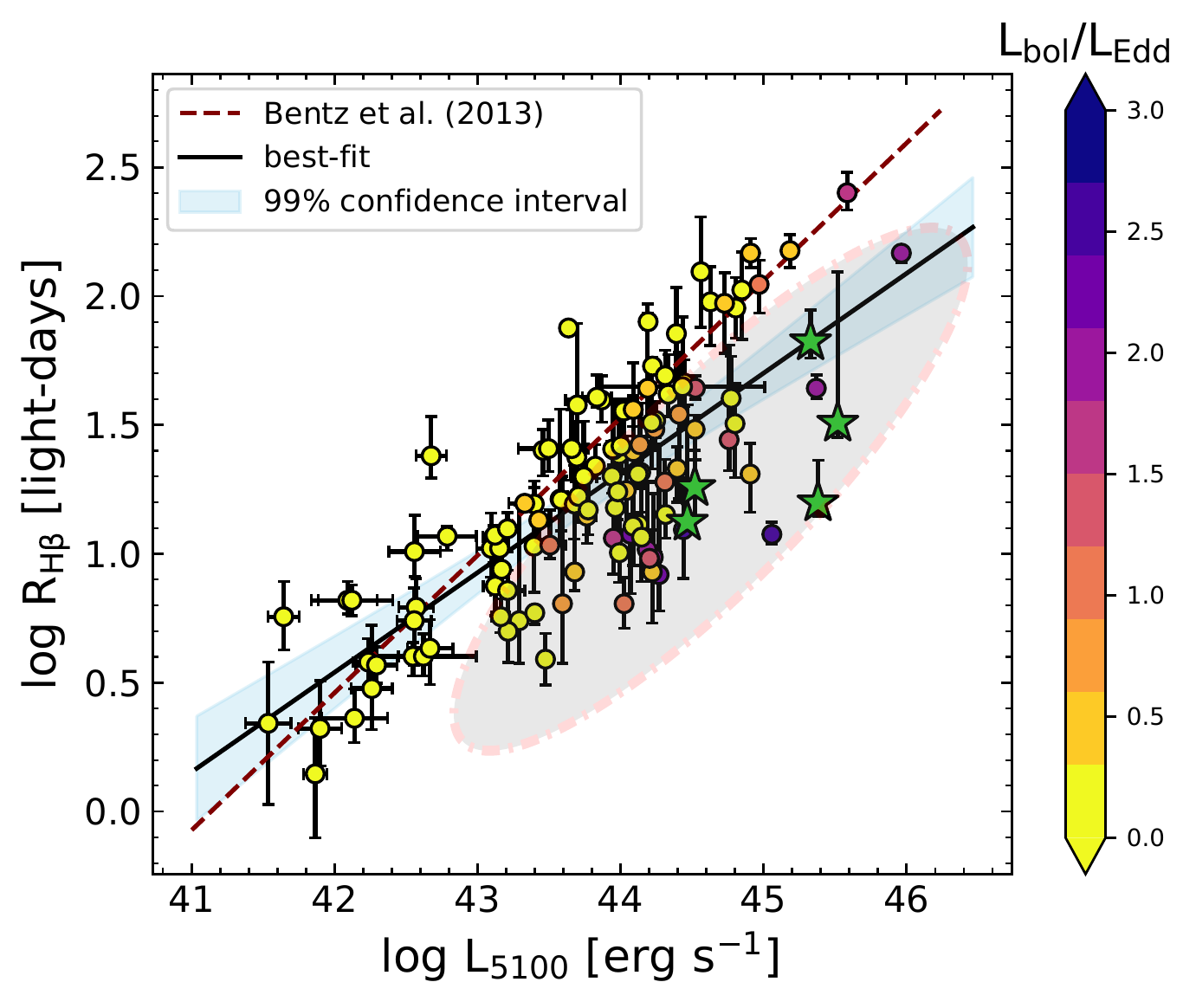}
    \includegraphics[width=0.495\textwidth, height=0.42\textwidth]{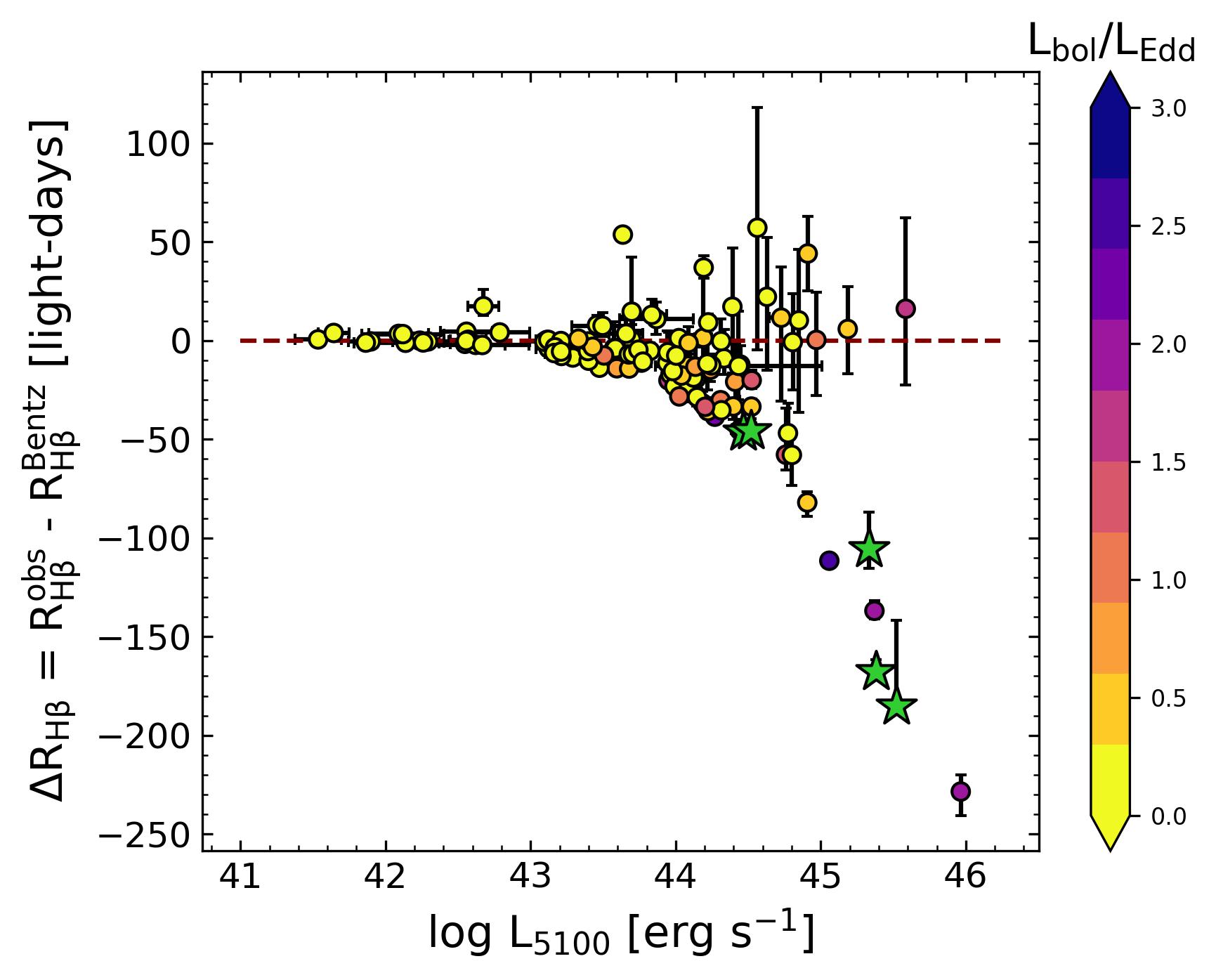}

    \caption{(Left:) BLR radius (of \hb{} emitting region) versus the AGN monochromatic luminosity at 5100\AA. The sources are coloured with respect to their Eddington ratios (L$_{\rm bol}$/L$_{\rm Edd}$). The dashed line shows the classical relation from \cite{bentz13}. The linear best-fit relation (black solid line) for the sources has the form: log R$_{\rm H\beta}$ = 0.387$\times$($\log\;{\rm L_{5100}}$) - 15.702, with a Spearman's correlation coefficient $\rho\approx$ 0.733 and p-value $\approx$\ 2.733$\times 10^{-21}$. The shaded region in light blue marks the 99\% confidence interval about the linear best-fit relation. The shaded ellipse highlights the sources with relatively high Eddington ratio values that deviate away from the classical relation, i.e., towards shorter BLR radii. Sources with \lledd $>$ 3 are highlighted using star symbols in green. Data are from \cite{martinez-aldama2019} which compiled the observational data for the 117 AGNs that consists of 48 sources previously monitored by \citet{bentz2009,bentz2014}, \citet{barth2013}, \citet{pei2014}, \citet{bentz2016}, and \citet{fausnaugh2017}, 25 super-Eddington sources of the SEAMBH project \citep[Super-Eddington Accreting Massive Black Holes,][]{Wang2014_seambh,du2015,hu15,dupu2016a,dupu2018}, 44 sources from the SDSS-RM \citep{grier17} sample and the recent monitoring for NGC5548 \citep{lu2016} and 3C273 \citep{zhang2019}. (Right:) The difference between the observed time delay (as shown with the data points in the left panel) and the time delay predicted by the classical relation from \cite{bentz13}. Here, the dashed line represents the null difference between the two delays.}
    \label{fig:rl-relation}
\end{figure}

An important result of the reverberation mapping studies is from the empirical power-law relation between the BLR radius (or light travel time delay) and the luminosity, \rblr{} $ \approx c \tau\ \propto $ \lopt{} $^\alpha$, where $\tau$\ is the time delay in response to continuum variation of a suitable line, mostly \hb\footnote{Here, the relation assumes the BLR radius for the H$\beta$ and the nearest continuum luminosity at 5100\AA}.  \citet{bentz13} found a best-fit for a sample of 41 AGNs covering four orders of magnitude in luminosity with a power-law slope value, $\alpha$=0.533$^{+0.035}_{-0.033}$, very close to the theoretical value, $\alpha = 0.5$\ needed to preserve spectral similarity \citep{wills1985, davidson_1977, osterbrock_ferland06}. This function is shown using a dashed line in the left panel of Figure \ref{fig:rl-relation}. One can then combine the \rblr{}-\lopt{} relation with the line widths for the broad emission lines estimated from single/multi-epoch spectroscopy to estimate the black hole masses which makes it especially useful for large statistical surveys of sources throughout cosmic history \citep{vester06, s11}.  

\begin{figure}[!htb]
    \centering
    \includegraphics[width=\textwidth]{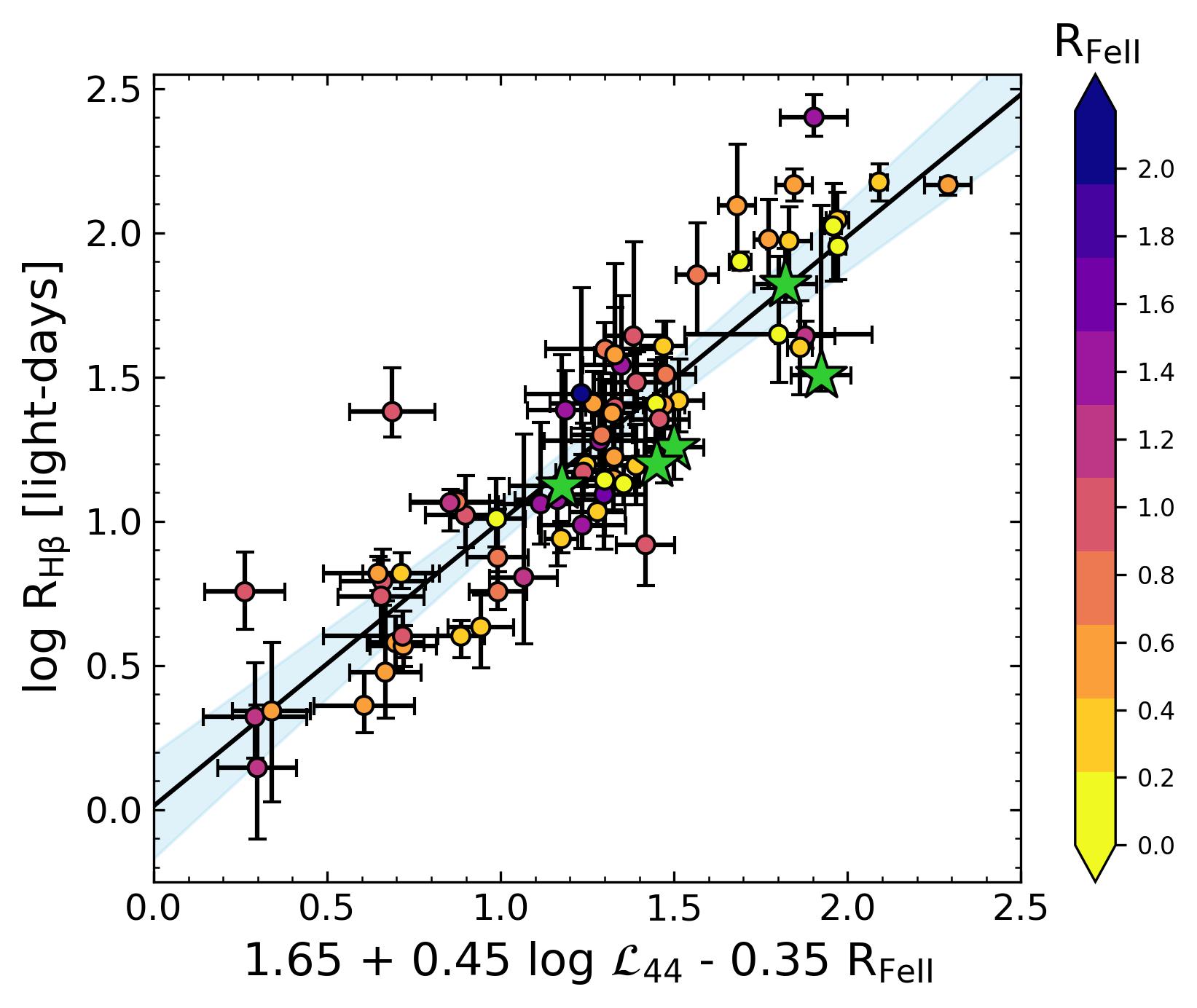}
    
    \caption{An abridged version of Figure 5 in \citet{du2019} showing the new \rblr{} - \lopt{} relation with the inclusion of the \rfe{}. The sample contains 75 reverberation-mapped sources which have \rfe{} estimated from their respective spectrum. The best-fit relation is shown in a black solid line with $\rho$ = 0.894 and p-value = 3.466$\times 10^{-27}$. The corresponding shaded region (in light blue) highlights the 99\% confidence interval about the linear best-fit relation with an effective scatter of $\sim$0.196 dex. Here, $\mathcal{L}_{44}$ represents the \lopt{} normalized by 10$^{44}$ erg s$^{-1}$ consistent with the formalism of \citet{du2019}. Sources with \lledd $>$ 3 are highlighted using star symbols in green as shown in Figure \ref{fig:rl-relation}.}
    \label{fig:rl-relation_alt}
\end{figure}

Recent observations have led to populate the \rblr{}-\lopt{} observational space and take the total count over 100, especially the sources monitored under the SEAMBH project (Super-Eddington Accreting Massive Black Holes, \citealt{dupu2014, Wang2014_seambh, hu15, du2015, dupu2016a, dupu2018}), and from the SDSS-RM campaigns \citep{grier17, Shen_2019ApJS..241...34S}. But this has introduced us to a new challenge - the inherent dispersion in the \rblr{}-\lopt{} relation after the introduction of these new sources. The left panel of Figure \ref{fig:rl-relation} is an abridged version from \citet{martinez-aldama2019, Panda_2021PhDT........22P} where the \rblr{}-\lopt{} observational space for 117 reverberations mapped AGNs is shown. The sources are coloured with respect to their Eddington ratios (\lbol{}/\ledd{}). The best-fit relation for this sample is, log R$_{\rm H\beta}$ = 0.387$\times$($\log\;{\rm L_{5100}}$) - 15.702, with a Spearman's correlation coefficient ($\rho$) = 0.733 and p-value = 2.733$\times 10^{-21}$, thus making the overall slope of the relation much shallower than obtained from the previous studies by \citet{bentz13} and bringing the validity of the empirical \rblr{}-\lopt{} relation into question. But interestingly, the sources that eventually led to the increase in the scatter in the relation show a trend with  the Eddington ratio - the larger the dispersion of a source from the empirical \rblr{}-\lopt{} relation, the higher its Eddington ratio! In \citet{martinez-aldama2019}, we found that this dispersion can be accounted for in the standard \rblr{}-\lopt{} relation with an added dependence on the Eddington ratio (\lbol{}/\ledd{}). This is highlighted in the right panel of Figure \ref{fig:rl-relation}, where the difference between the observed time delays and that estimated from the empirical \rblr{}-\lopt{} relation \citep[i.e., $\Delta$\rblr{},][]{bentz13}, for the sources shown in the left panel, are plotted against their respective \lopt{}. One can appreciate the drop in the $\Delta$\rblr{} value, especially for sources with high Eddington ratios indicating that sources with shorter `observed' time lag especially at the high luminosity end of the scaling relation exhibit high accretion rates. In other words, these sources are expected to host SMBHs with low BH masses \citep{dupu2014,dupu2016a}. Although we note that there are some sources which, albeit being of high Eddington nature, are observed to have \rblr{} sizes comparable to the predicted value from \citet{bentz13}. \citet{du2019} exploited this further in their work and realized that with an additional correction term, the relation can be reverted back to the original relation with a slope $\sim$0.5. This additional correction term is an observational parameter, the relative strength between the optical \feii{} emission and the corresponding H$\beta$ emission (\rfe{}) that has been shown in earlier studies to be a reliable observational proxy for the Eddington ratio \citep{sulentic2000,marzianietal01,sh14, mar18, panda19b, du2019, martinez-aldama_2021} which we touched upon in earlier sections. The relation takes the form  \citep{du2019}: 

\begin{equation}
\log\left(\frac{{\rm R_{BLR}}}{1\;\mathrm{light-day}}\right) = \kappa + \alpha{}\log\left(\frac{{\rm L_{5100}}}{10^{44}\, \mathrm{erg\ s^{-1}}}\right) + \gamma{}{\rm R_{FeII}}
\label{eq:rcorr}
\end{equation}

where the flux (F) can be independently estimated from the observed AGN spectrum for a given source. {With this correction in terms of an observable quantity, i.e., \rfe{}, we avoid the circularity problem which was present when the correction was explicitly made in terms of the Eddington ratio \citep{martinez-aldama2019}, } and can have a robust estimate of the luminosity distance (D$_{\rm L}$). This then allows us to construct a Hubble diagram using quasars where we know their luminosity distance and their redshifts, and test the validity of the standard cosmological model and their alternatives \citep[see e.g.][]{Haas2011, watson2011, czerny2013, Czerny_etal_2019, czernyetal21, zajacek2021, khadka2021, khadka2022}. {Super-Eddington accreting sources are expected to be preferentially selected with increasing redshift  {in flux-limited sample}\citep{sulenticetal14}. {A combination of this selection bias and intrinsic Eddington ratio evolution  \citep[e.g.,][]{cavalierevittorini00,hopkinsetal06} makes for high redshift quasar spectra often resembling lower-$z$\ extreme Population A spectra (an effect predicted by \citealt{sulentic2000}).} Their inclusion is thus}  vital in extending the \rblr{}-\lopt{} relation to higher luminosity regime as shown in Figure \ref{fig:rl-relation_alt}.
where, $\kappa$=1.65$\pm$0.06, $\alpha$=0.45$\pm$0.03, and $\gamma$=-0.35$\pm$0.08. Clearly, the introduction of the \rfe{} term and for sources with strong \feii{} emission, is able to account for their shorter time-lags and hence, smaller \rblr{} sizes.   In addition, we are able to recover the slope ($\alpha$) closer to the theoretical predictions, i.e., 0.5 \citep{davidson_1977, Davidson_Netzer_1979} and hence, we can safely use the modified \rblr{}-\lopt{} relation (including the correction term wrt \rfe{}) to independently estimate the monochromatic luminosity for a given source, and couple the information from the flux obtained from direct observations, to eventually estimate the luminosity distance  to the said source: 

\begin{equation}
{\rm D_L} = \sqrt{{\rm L_{5100}}/4\pi {\rm F}} \propto \frac{c\tau}{\sqrt{4\pi F}}
\label{eq:dl}
\end{equation}

where the flux (F) can be independently estimated from the observed AGN spectrum for a given source. In this way, we can avoid circularity and can have a robust estimate of the luminosity distance. 

\subsection{Anisotropic radiation from the accretion disk}\label{sec3}


An  equally important aspect in this regard is the ionizing continuum produced by the central engine.
The characterization of the ionizing SED that comes from regions closer than the BLR is important for our study of the emission lines. 
From the photo-ionization point of view, this fraction of the broad-band SED is closely related to the number of ionizing photons that eventually leads to the line production that has permitted the estimation of a photoionization radius of the BLR \citep{wandel99, negrete2014, martinez-aldamaetal15, panda_cafe2, Panda_2022FrASS...950409P}.

We tested the variation in the low-ionization part of the BLR by accounting for the changes in the shape of the ionizing continuum (the SED) and the location of the \hb{}-emitting BLR from the central ionizing source (or \rblr{}) from the reverberation mapping, in the context of Main Sequence of Quasars \citep{Panda_2022FrASS...950409P}. In this and previous work  \citep{panda_cafe2}, we have found that in order to estimate the correct physical conditions for these low-ionization lines emitting regions in the BLR, it is not sufficient to only retrieve the flux ratios (e.g., \rfe{}) but to also have an agreement with the corresponding modelled and observed line strengths (or line equivalent widths, EWs). Compared to the results that are directly obtained from the photoionization theory, {these new results highlight the shift in the overall location of the line-emitting \rblr{} - in terms of the ionization parameter  U  and the local cloud density (n$_{\rm H}$) recovered from the analysis towards lower values (by up to 2 dexes) compared to the \rblr{} values estimated from the photoionization theory.} This brings the modelled location in agreement with the reverberation mapping results, especially for the high-accreting NLS1s which show shorter time-lags/smaller emitting regions. A corollary result is that to retrieve such physical conditions, the BLR should ``see'' a different, filtered SED with only a fraction of the total ionizing photon flux. This analysis was performed on selected sources with readily available  broad-band SEDs and archival spectroscopic measurements. In addition, we  assumed source-specific metallicities that were derived using the UV diagnostic lines from earlier studies \citep[see e.g.,][]{Marziani_etal_2022}. There is a need to extend this analysis to a larger number of reverberation-mapped sources. We, therefore, need synchronous multi-wavelength observations to build robust SEDs that can be used to confirm this scenario. Also, there is a need to bring together a global picture where a combined analysis of the UV and optical emitting regions can be put together -- to allow us to gauge the salient differences in the low- and high-ionization line emitting regions. 


\begin{figure}[!htb]
\centering
\includegraphics[width=1\textwidth]{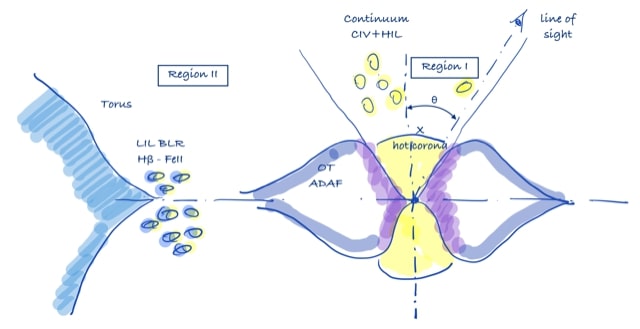}
\caption{
Schematic view of the inner sub-parsec region around the SMBH for a high accreting AGN. {Region I is exposed to the continuum emitted from the hottest part of the optically thick (OT) and geometrically thick advection-dominated accretion flow (ADAF). The observer sees inside of this region if the line-of-sight  is inclined by $\lesssim 30$\ degrees from the disk axis. Region II refers to the region shielded from the hottest region and  exposed to a colder continuum emitted from the ADAF at angles $\gtrsim \theta$. Abridged version from \cite{wang14}; not drawn to scale.  }}
\label{fig:my_label}
\end{figure}

\citet{wang14} derived the analytical solutions (steady-state) for the structure of ``slim'' accretion discs from sub-Eddington accretion rates to extremely high, super-Eddington rates. They notice the appearance of a funnel-like structure very close to the SMBH at super-Eddington rates and attribute this feature to the puffing up of the inner accretion disk by radiation pressure,  wherein  the \cite{ss73} prescription for a  geometrically thin, optically thick accretion disk does not hold \citep[][]{abramowicz88, sadowski2011, wang14}. We show an illustration of this scenario in   Figure \ref{fig:my_label}. Such modiﬁcations to the disk structure strongly affect the overall anisotropic emission of ionizing photons from the disk in addition to just inclination effects   that arise due to the axisymmetric nature of these systems. Therefore, with a rise in accretion rates, a continuum anisotropy needs to be accounted for. The anisotropy  then leads to the shrinking in the position of the BLR -- exposed to a lower continuum flux than the observer (Figure \ref{fig:my_label}) -- that brings the modelled location in agreement with the observed estimates from the reverberation mapping campaigns \citep[see][for more details]{panda_cafe2}.

\begin{figure}[!htb]
    \centering
    \includegraphics[width=0.495\textwidth, height=0.4\textwidth]{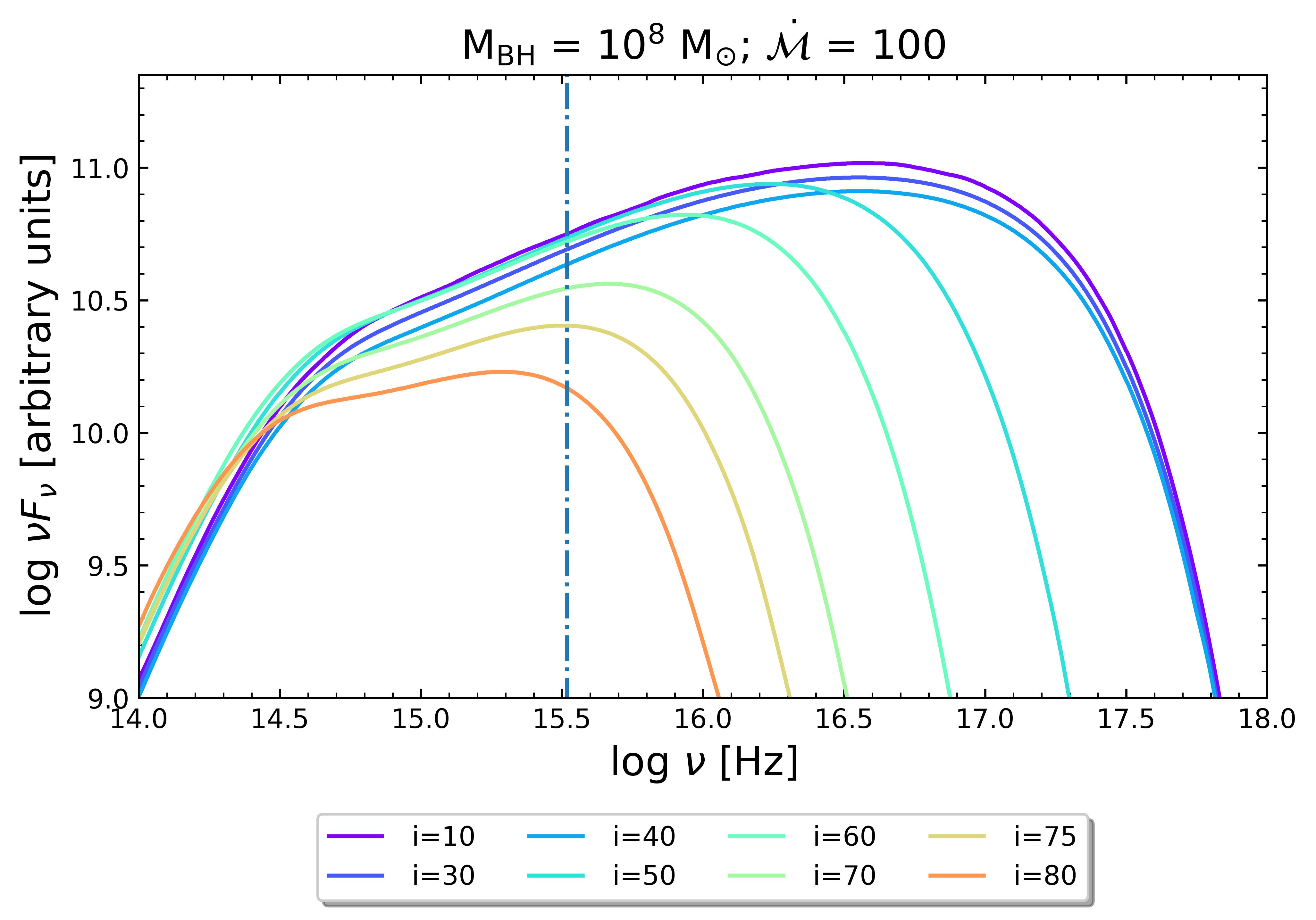}
    \includegraphics[width=0.495\textwidth, height=0.4\textwidth]{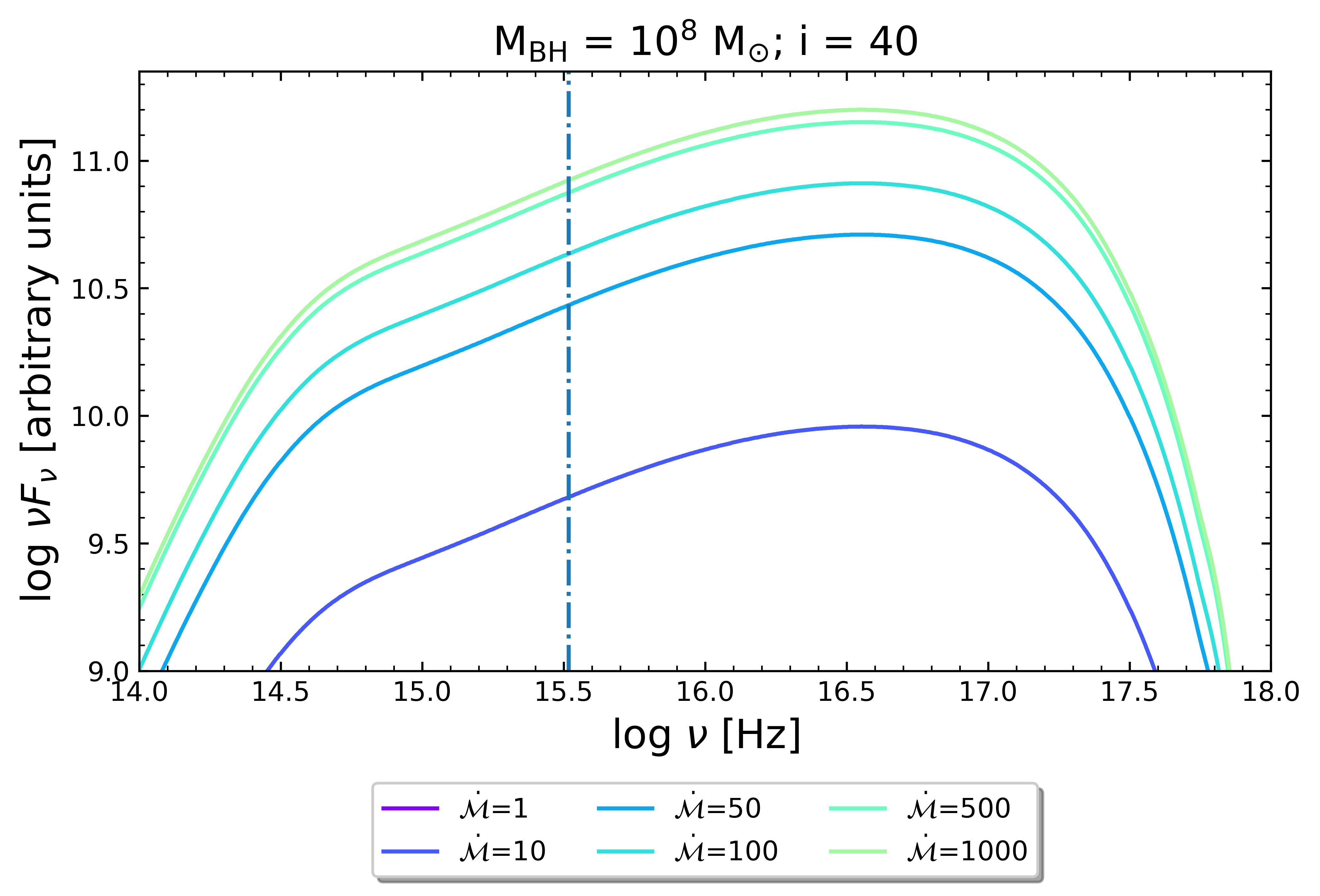}
    \caption{Spectral energy distributions (SEDs) obtained for slim accretion disks for a representative black hole mass, \mbh{} = 10$^8$ \msun{}. LEFT: SEDs are shown for a range of viewing angle cases for a representative dimensionless accretion rate, $\mathcal{\dot{M}}$ = 100. RIGHT: SEDs are shown for a range of $\mathcal{\dot{M}}$ for a representative viewing angle, \textit{i} = 40$^{\circ}$. In both panels, the vertical dash-dotted line marks the 1 Rydberg threshold.}
    \label{fig:slim-disks}
\end{figure}    

\begin{table}[!htb]
\centering
\caption{Fraction of ionizing  continuum flux  (in \%) of the slim disk SED  shown in the left panel of Figure \ref{fig:slim-disks} (relative to the case with \textit{i} = 10$^{\circ}$)}
\label{tab:table1}
\begin{tabular}{cc}\hline
\textit{i} & ratio (\%) \\ \hline
10$^{\circ}$    & 100.00   \\
30$^{\circ}$    & 87.91    \\
40$^{\circ}$    & 77.92    \\
50$^{\circ}$    & 26.11    \\
60$^{\circ}$    & 7.95     \\
70$^{\circ}$    & 1.91     \\
75$^{\circ}$    & 0.78     \\
80$^{\circ}$    & 0.23     \\ \hline
\end{tabular}%
\end{table}

\begin{table}[!htb]
\centering
\caption{Fraction of ionizing continuum flux (in \%) under the slim disk SEDs shown in the right panel of Figure \ref{fig:slim-disks} (normalized to the case with $\mathcal{\dot{M}}$ = 1000)}
\label{tab:table2}
\begin{tabular}{ccc}\hline
\multicolumn{1}{c}{$\mathcal{\dot{M}}$} & \multicolumn{1}{c}{ratio (\%)} & \multicolumn{1}{c}{ratio $\mathcal{\dot{M}_\mathrm{i}}$  / ratio $\mathcal{\dot{M}}_{i-1}$} \\ \hline
1                         & 0.44  & \ldots                       \\
10                        & 5.71    & 12.98                   \\
50                        & 32.40   & 5.67                   \\
100                       & 51.49   & 1.59                   \\
500                       & 89.48   & 1.74                   \\
1000                      & 100.00  & 1.12                   \\ \hline
\end{tabular}%
\end{table}

The left panel in Figure \ref{fig:slim-disks} shows the slim-disk SEDs (Jian-Min Wang, priv. comm.) for a representative BH mass of 10$^{8}$ \msun{}, accreting at $\mathcal{\dot{M}}$ = 100 for a range of viewing angles\footnote{This is the dimensionless accretion rate introduced by \cite{Wang2014_seambh}: $\dot{\mathcal{M}} = \dot{M}/\dot{M}_\mathrm{Edd}$, with $\dot{M}_\mathrm{Edd} = L/c^2$, and $\dot{M}$\ the  mass accretion rate. In \cite{Panda_2022FrASS...950409P} we provide a analytical form to convert $\mathcal{\dot{M}}$ to Eddington ratio (\lbol{}/\ledd{}, see equation 13 in their paper). This relation additionally depends on the BH mass and the bolometric correction. For a BH mass of 10$^{8}$ \msun{} with $\mathcal{\dot{M}}$ = 100, for a \lopt{} = 10$^{45}$ erg s$^{-1}$, the Eddington ratio is $\sim$0.1.}. The right panel shows the distribution of slim-disk SEDs as a function of $\mathcal{\dot{M}}$ for a representative BH mass of 10$^{8}$ \msun{} observed at a viewing angle (\textit{i} = 40$^{\circ}$). We also report the relative area under the SEDs shown in Figure \ref{fig:slim-disks}. These values are tabulated in Tables \ref{tab:table1} and \ref{tab:table2} corresponding to the left and right panels of Figure \ref{fig:slim-disks}, respectively. We estimate the area under the SEDs accounting only for the fluxes corresponding to a frequency $\geq$ 1 Rydberg\footnote{1 Rydberg $\approx$ 3.29$\times 10^{15}$ Hz}. We then compute the relative area (a) with respect to the SED case with the viewing angle, \textit{i} = 10$^{\circ}$ (Table \ref{tab:table1} and left panel of Figure \ref{fig:slim-disks}); and (b) with respect to the SED case with the dimensionless accretion rate, $\mathcal{\dot{M}}$ = 1000 (Table \ref{tab:table2} and right panel of Figure \ref{fig:slim-disks}).
From the left panel of Figure \ref{fig:slim-disks}, we can notice that going from the SED viewed at \textit{i} = 10$^{\circ}$ to 80$^{\circ}$, keeping the BH mass and accretion rate constant, the extended region II  receives only a very small fraction of the actual {\em ionizing} photon flux (only 0.23\%), meaning almost all of the ionizing photons (99.77\%) never make it to the BLR. This $\sim$2 dex reduction in the photon flux results in an equal reduction in the ionization parameter (U) which was confirmed already in \citet{panda_cafe2}. On the other hand, changing the accretion rate, going from $\mathcal{\dot{M}}$ = 1 to 1000 and keeping the BH mass and viewing angle constant (right panel of Figure \ref{fig:slim-disks}),  {we find that an accretion rate $\mathcal{\dot{M}}$ = 1 relates to only a 0.44\% of the total photon flux, } also a factor $\gtrsim 100$\ from the case $\mathcal{\dot{M}}$ = 1000. There is however a fundamental difference: whereas changing $\mathcal{\dot{M}}$\ produces an almost  self-similar shift of the SED, a change in the viewing angle produces the change by a factor $\gtrsim 100$\ in the {\em ionizing flux}, while the optical flux changes by a factor $\approx a few$\ (Figure \ref{fig:slim-disks}). This indirectly confirms the results of previous works that have been pointing to the main sequence drive being the Eddington ratio convolved with the effect of  orientation \citep{marzianietal01, sh14, sun15, mar18, panda19b}. The shape of the SED thus plays an important role in explaining the trends in the quasar main sequence wherein the information of the fundamental BH parameters -- BH mass, Eddington ratio, orientation and the BH spin -- are embedded \citep[see][for more details]{Panda_2021PhDT........22P}.

\begin{figure}[!htb]
\centering
\includegraphics[width=1\textwidth]{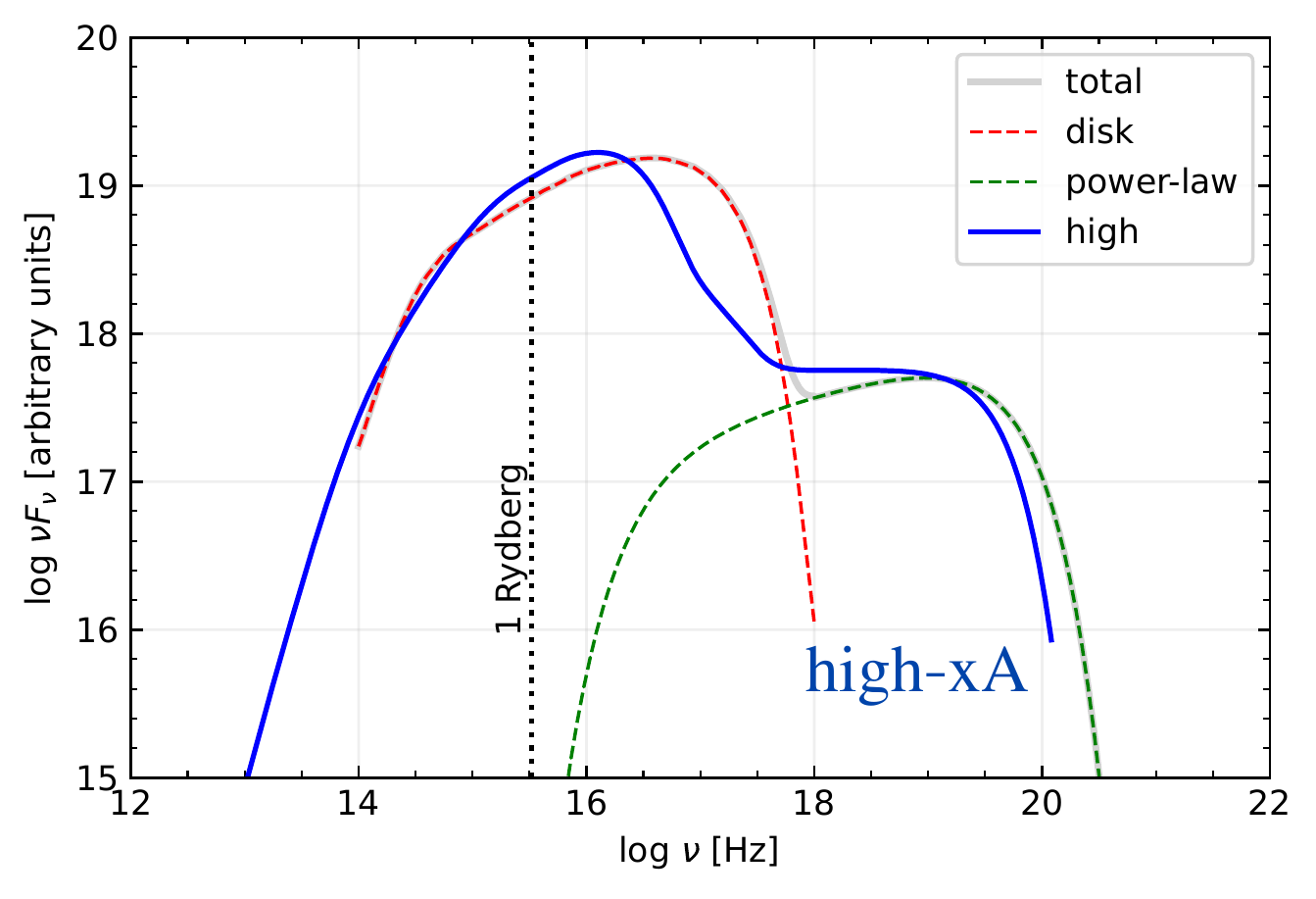}
\caption{
A simplified model for the "high" SED of \citet{ferland2020} using the $\mathcal{\dot{M}} = 500$\ SED from \citet{wang14} and an X-ray emitting corona (power law with exponential breaks). Note that the high-energy turnover at $\log \nu \sim 20$\ [Hz] is actually poorly known, and in the most extreme case, the hard X-ray SED may show no flattening and no break (``highest" case; magenta line in Figure \ref{fig:sedax}).}
\label{fig:xamod}
\end{figure}

\begin{figure}[!htb]
\centering
\includegraphics[width=1\textwidth]{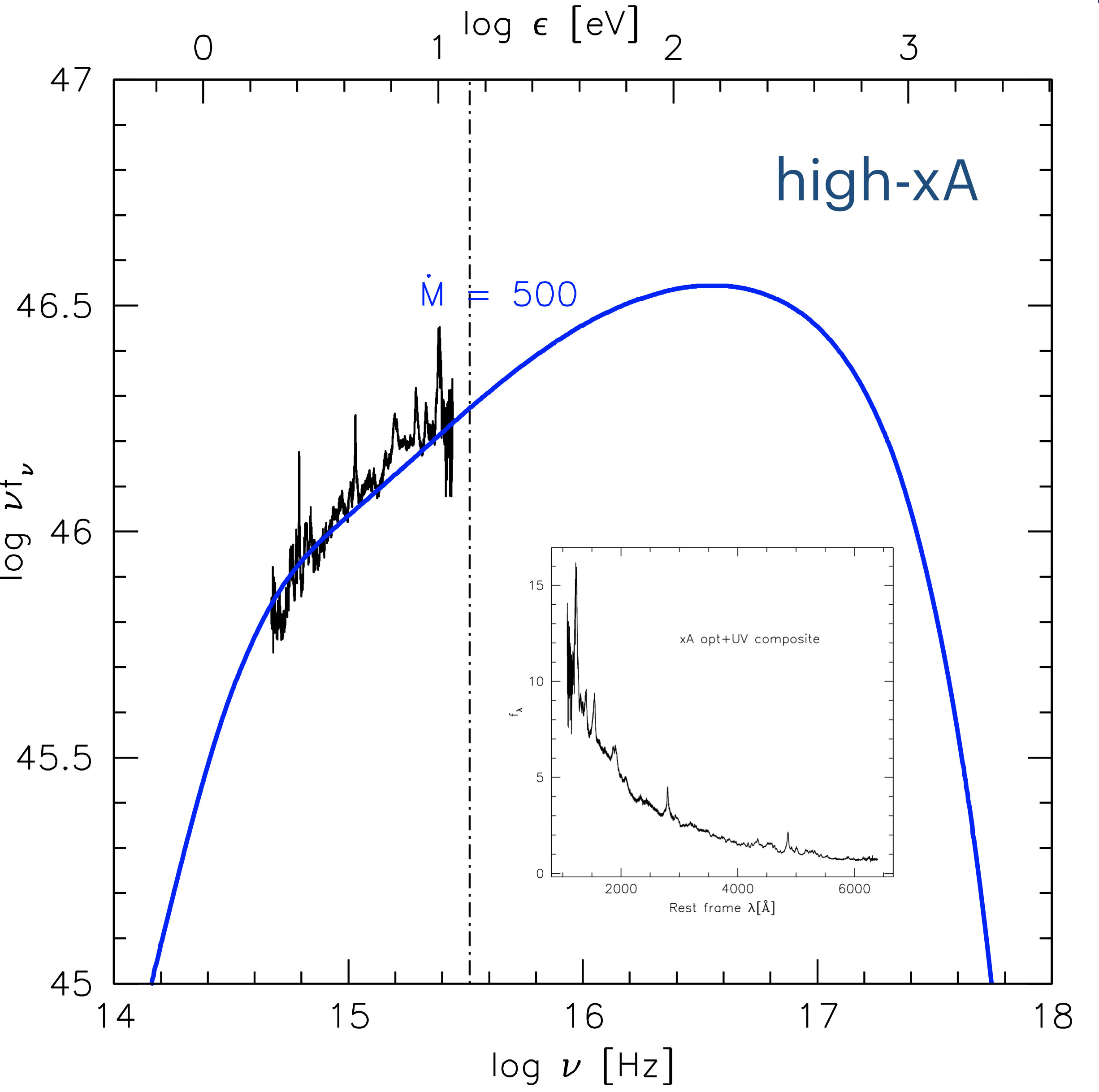}
\caption{
Composite spectrum for xA from \citet{marzianietal13}, with the \citet{wang14} SED superimposed for $\mathcal{\dot{M}} = 500$. The inset shows the same spectrum as a function of wavelength. }
\label{fig:xaobs}
\end{figure}

{In Figure \ref{fig:xamod}, we highlight a representative fit (in grey) to the high SED case (in blue) from \citet{ferland2020} utilizing a two-component model, i.e., a slim accretion disk that represents the thermal component (in dashed red) along with a hot Comptonized component (in dashed green). Here, the slim accretion disk SED is modelled for a face-on viewing angle, \textit{i} = 10$^{\circ}$ for a $\mathcal{\dot{M}}$ = 500 for a BH mass of 10$^{8}$ \msun{}. The X-ray to optical-UV normalization is set by a spectral index ($\alpha_{\rm ox}$ = -1.47) and the high-energy cutoff is assumed to be $\sim$2.5 keV with a slope ($\alpha_{\rm x}$ = -0.79) to match the exponential drop in the observed SED. The assumed values for these parameters are considered from \citet{jin12a, jin12b} who were the first to carry out a broadband analysis on these composite SEDs. This instance of the slim accretion disk SED is a good fit for xA as shown in Figure \ref{fig:xaobs} where the latter is represented by a composite spectrum made using the optical-UV spectral observations for xA sources from \citet{marzianietal13}. In our forthcoming work, we will incorporate these slim disk SEDs into our photoionization modelling setup and recover the trends for the low- and high-ionization emission lines, their relative strengths (e.g., \rfe{}) and the EWs, concerning these fundamental BH parameters, concentrating on the xA.}

Table \ref{tab:table2} shows that the fraction of ionizing photons at a very high accretion rate tends to saturate, with only a 10\%\ increasing from the doubling of the accretion rate, from $\mathcal{\dot{M}} = 500$\ to 1000. At such $\mathcal{\dot{M}}$ the Eddington ratio should converge toward a limiting value of $\mathcal{O}(1)$. The product $n_\mathrm{H}U$\ is also little affected by   changes in SED in the cases  shown in Figure \ref{fig:sedax}. The factors $\mathcal{S}$\ and $\mathcal{P}$\ are expected to be stable,\footnote{The average frequency of the ionizing continuum changes by just 5\%\ passing from the \citet{mathewsferland87} to \citet{marzianisulentic14}: from $<h\nu> \approx 3.03$\ to $<h\nu> \approx 3.17$\ Ryd. } although their actual dispersion and systematics for sources selected as radiating closer to the Eddington limit should be further investigated through dedicated observations.  In other words,  even if anisotropy effects in line widths are strong, anisotropy in continuum emission and differences in SEDs  might not be so strong as to compromise an application to the cosmology of Eq. \ref{eq:lvir} that is -- we stress it  -- generally valid for all AGNs but in practice  exploitable for high accretors only. Preliminary applications to cosmology of Eq. \ref{eq:lvir} have been encouraging \citep{marzianisulentic14,marzianisulentic14a,marzianietal19,czernyetal21,marzianietal21,marzianietal21a}.


\section{Concluding remarks}

We have reviewed some basic aspects of luminous type-1 AGN (Section \ref{sec1}) to introduce highly accreting sources with special attention to their SEDs (Section \ref{sec2}), focusing on the most relevant aspects (such as anisotropy, Sect. \ref{sec3}) in the context of the  possible application to the measurements of the cosmological parameters (Sections \ref{sec2} and \ref{cosmoxa}). We have stressed that Population A includes NLS1s but that only a fraction of the sources can be considered highly accreting, as a defining criterion is \rfe $\gtrsim 1$, and this criterion is not met by all Population A and NLS1 sources. On the converse, a criterion based on line width implies a dependence on black hole mass (and hence luminosity in flux-limited samples), viewing angle, Eddington ratio, and yields a selection that is, unavoidably, sample dependent \citep{marzianietal18, panda19b}.

\begin{figure}[!htb]
    \centering
    \includegraphics[width=\textwidth]{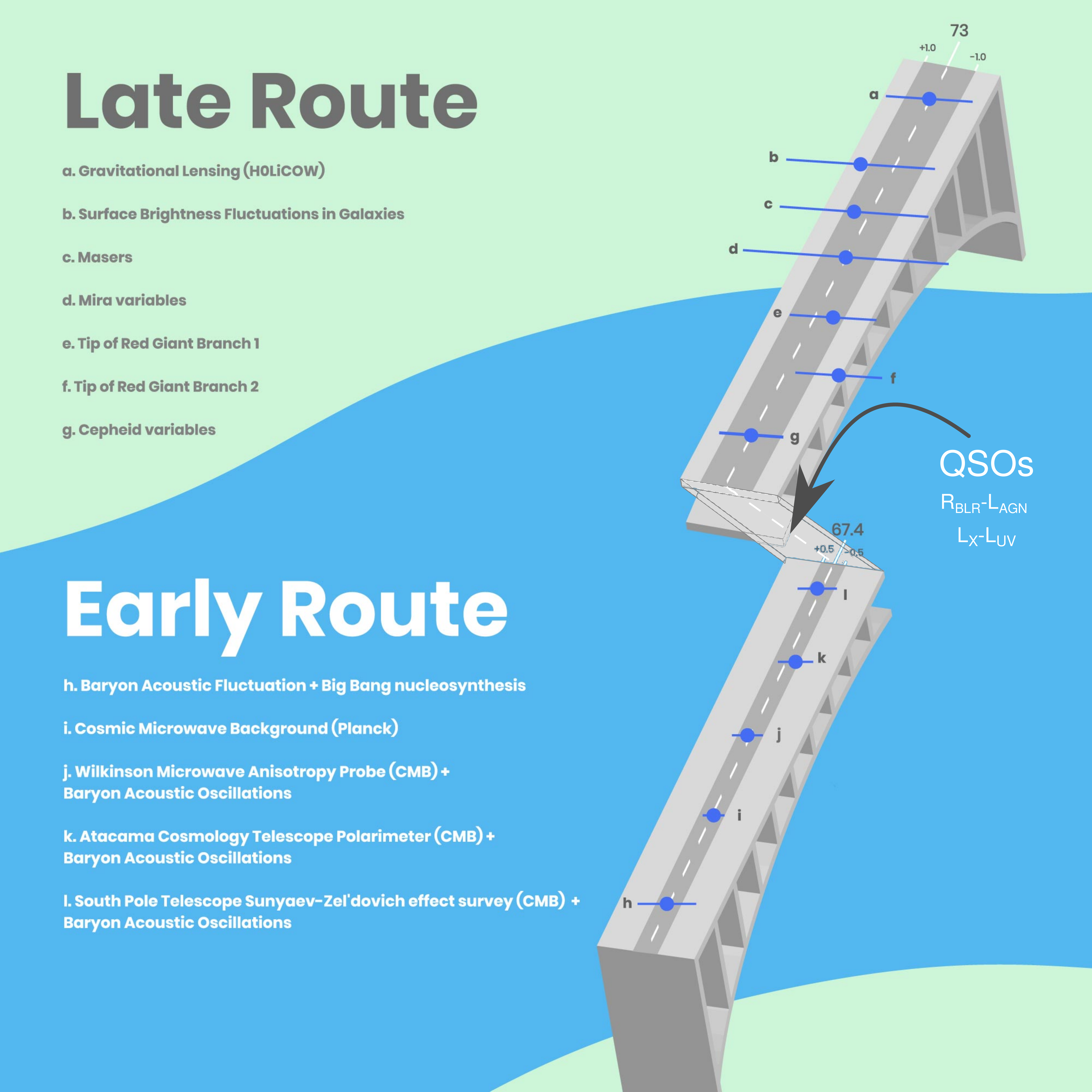}
    \caption{This graphic lists the variety of techniques that have been used to measure the expansion rate of the universe, known as the Hubble constant (${\rm H_0}$). One set of observations looked at the very early universe (or the early route, shown in the bottom half of the graphic) and the second set of observation strategies analyzed the universe’s expansion in the local universe (or the late route, shown in the upper half of the graphic). The letters corresponding to each technique are plotted on the bridge on the right. The location of each dot on the bridge road represents the measured value of the ${\rm H_0}$, while the length of the associated bar shows the estimated amount of uncertainty in the measurements. The combined average from the seven methods from the late route yields a ${\rm H_0}$ value of 73 km s$^{-1}$ Mpc$^{-1}$. This number is at odds with the combined value of the techniques used to calculate the universe’s expansion rate from the early route. Their combined value for the ${\rm H_0}$ is 67.4 km s$^{-1}$ Mpc$^{-1}$. Abridged version. Original graphic credit: NASA, ESA, and A. James (STScI).}
    \label{fig:cosmic-ladder}
\end{figure}

Looking at the bigger picture,    we can construct the Hubble diagram with the luminosity distances using Eq. \ref{eq:dl} corrected according to Eq. \ref{eq:rcorr} and the corresponding redshifts for each source. The key here is to have the measurement of the time-delay (e.g., from Eq. \ref{eq:rcorr}) and the AGN monochromatic flux using single epoch spectra that allow us to estimate the luminosity distances regardless of accretion properties. Hence, very large samples of reverberation-mapped AGNs can be used as cosmological candles \citep[][]{collier1999, elvis2002,horne2003,panda_2019_frontiers, khadka2022}. This may further allow us to study the evolution of the cosmological parameters as a function of the redshift allowing for the reconciliation of the Hubble tension - the disparity between the measured value of the Hubble constant in the local and the early Universe (see Figure \ref{fig:cosmic-ladder}).

There will be an immense potential for the ideas and results presented in this work in the near future, serving as test-beds for the vast number of AGNs that will be explored with the ongoing and upcoming ground-based 10-metre-class \citep[e.g. Maunakea Spectroscopic Explorer,][]{2019BAAS...51g.126M} and 40 metre-class \citep[e.g. The European Extremely Large Telescope,][]{2015arXiv150104726E} telescopes; and space-based missions such as the JWST \citep{2006SSRv..123..485G, 2022A&A...661A..80J} and the Nancy Grace Roman Space Telescope \citep{2013arXiv1305.5422S}. Increased availability of high-quality, multi-wavelength photometric, spectroscopic and interferometric measurements extending to higher redshifts is a necessity to help develop our ever-growing theoretical understanding of how these massive, energetic cosmic sources work and evolve as well as how they might be exploited for cosmic distance estimates.

\section*{Conflict of Interest Statement}
The author declares that the research was conducted in the absence of any commercial or financial relationships that could be construed as a potential conflict of interest.

\section*{Author Contributions}
All authors listed have made a substantial, direct, and intellectual
contribution to the work and approved it for publication.

\section*{Funding}

\section*{Acknowledgments}
SP acknowledges the Conselho Nacional de Desenvolvimento Científico e Tecnológico (CNPq) Fellowship (164753/2020-6). SP further acknowledges the organizers of the SPIG - 31$^{\rm st}$ Summer School and International Symposium on the Physics of Ionized Gases, held between 05$^{\rm th}$ - 09$^{\rm th}$ September 2022, for allowing presenting our work as an invited contribution. We are grateful to Prof. Jian-Min Wang for providing his slim disk SED models and to Prof. Bo\.zena Czerny for fruitful discussions.

\section*{Data Availability Statement}
All data incorporated in this work can be made available upon request to the authors.


\bibliographystyle{frontiersinSCNS_ENG_HUMS} 
\bibliography{sn-bibliography}


\end{document}